\documentclass[twocolumn,prl,showpacs,preprintnumbers]{revtex4}
\usepackage[dvips]{graphicx}
\usepackage[usenames]{color} 
\usepackage{amsfonts,amsbsy}

\begin{document}

\newcommand{\ada}{adiabatic approximation}
\newcommand{\adt}{adiabatic theorem}
\setlength{\unitlength}{1cm}

\newcommand\bra[1]{\left\langle#1\right|}
\newcommand\ket[1]{\left|#1\right\rangle}
\newcommand\eqref[1]{(\ref{#1})}
\newcommand\beq{\begin{equation}}
\newcommand\eeq{\end{equation}}
\newcommand\bea{\begin{eqnarray}}
\newcommand\eea{\end{eqnarray}}
\newcommand\eps{\epsilon}
\newcommand\ltwid{\mathrel{
 \raise.3ex\hbox{$<$\kern-.75em\lower1ex\hbox{$\sim$}}}}
\newcommand\la{\lambda}
\newcommand\ga{{\gamma}}
\newcommand\om{{\omega}}
\newcommand\bfn{{\mathbf n}}
\newcommand\bfV{{\mathbf V}}
\newcommand{\apri}{{\em a priori\/}}
\newcommand{\apos}{{\em a posteriori\/}}
\newcommand{\bfsi}{{\boldsymbol \sigma}}

\preprint{UdeM-GPP-TH-07-159}

\title{Validity of the Adiabatic Approximation}
\author{R. MacKenzie, A. Morin-Duchesne, H. Paquette and J. Pinel}
\affiliation{Groupe de physique des particules, Universit\'e de Montr\'eal
C.P. 6128, Succ.~Centre-ville, Montr\'eal, QC H3C 3J7}

\begin{abstract}
We analyze the validity of the adiabatic approximation, and in particular the reliability of what has been called the ``standard criterion" for validity of this approximation. Recently, this criterion has been found to be insufficient. We will argue that the criterion is sufficient only when it agrees with the intuitive notion of slowness of evolution of the Hamiltonian. However, it can be insufficient in cases where the Hamiltonian varies rapidly but only by a small amount. We also emphasize the distinction between the adiabatic {\em theorem\/} and the adiabatic {\em approximation\/}, two quite different although closely related ideas.
\end{abstract}

\maketitle

\noindent{\em Introduction}~~~~~The {\adt} in quantum mechanics was developed when quantum mechanics was still in its infancy \cite{Born:1928}, and is a useful tool for analyzing problems where the Hamiltonian evolves slowly in time. It has come under some scrutiny recently, largely, we believe, due to a misunderstanding as to exactly what the theorem states, with many authors weighing in on both sides of the issue \cite{Marzlin:2004, Sarandy:2004, Pati:2004, Tong:2004, Ambainis:2004, Wu:2004, Tong:2005, Ye:2005, Vertesi:2005, Ma:2006, Duki:2006}.

We will attempt to identify this misunderstanding, and will make a distinction between the {\adt}, which is no more and no less than a mathematical theorem which can be (and has been) rigourously proven and which is beyond dispute, and the {\ada}, which is a statement about the conditions under which a system governed by a time-dependent Hamiltonian will to a good approximation have an evolution which is adiabatic. These issues will be clarified through an analysis of a concrete illustration of the inadequacy of what has been labelled (erroneously, we feel) the ``standard" criterion for the validity of the {\ada}.

Roughly speaking, the {\adt} states that a system whose dynamics is governed by a time-dependent Hamiltonian will, in the limit of infinitely slow evolution of the Hamiltonian, remain in the state which evolves from the initial state. More precisely, suppose that the Hamiltonian can be written
\beq
H(t)=E \hat H(t/\tau),
\label{zero}
\eeq
where $\hat H(s)$ is a dimensionless Hermitian operator depending smoothly on a dimensionless variable $s$ and $E$ is a positive constant with dimension of energy. $E$ and $\tau$ represent the energy scale and time scale of evolution of the Hamiltonian, respectively. We are interested in arriving at a given final Hamiltonian via slow evolution; mathematically, we wish to consider $\tau$ large with $s=t/\tau$ fixed. It is therefore useful to write the Schroedinger equation in terms of $s$ rather than $t$; defining $\eps=(E\tau)^{-1}$, we have
\beq
i\eps{d\over ds}\ket{\psi(s)}=\hat H(s)\ket{\psi(s)}.
\label{zeroa}
\eeq
We define instantaneous eigenstates and energies as the solutions of a family of time-independent Schroedinger equations parameterized by $s$, $\hat H(s)\ket{n(s)} =\hat E_n(s)\ket{n(s)}$, and choose the phases of the states so that $\langle n(s)|{\dot {n}(s)}\rangle=0$. Suppose that the system is initially in the state $\ket{0(0)}$ (not necessarily the ground state) and that the gap between $\hat E_0(s)$ and the other energy eigenvalues is bounded from below by unity. The {\adt} states that as $\eps\to0$, the solution to the time-dependent Schroedinger equation approaches 
\beq
\ket{\psi_{ad}(s)}=e^{-(i/ \eps) \int_0^s \hat E_0(s')ds'}\ket{0(s)},
\label{one}
\eeq
within an error of order $\eps$.

Closely related to but independent of the {\adt} is the following question: under what circumstances will a physical system governed by a time-dependent Hamiltonian and initially in an instantaneous eigenstate of the initial Hamiltonian evolve adiabatically, that is to say, when will the exact solution of the Schroedinger equation satisfy the {\ada}, $\ket{\psi(t)}\simeq\ket{\psi_{ad}(t)}$? One answer is if the conditions of the adiabatic theorem are satisfied. Yet circumstances other than these might also lead to adiabatic evolution, and it is obviously useful (and surprisingly nontrivial) to clarify such circumstances. 


\noindent {\em General considerations}~~~~~Let us return to the case of a Hamiltonian of the form given in \eqref{zero}. A perturbative calculation of $\ket{\psi(t)}$ can be given which is useful in adiabatic problems; an early classic reference is \cite{Messiah:1962}. In a more recent and perhaps more intuitive work \cite{mackenzie-2006-73}, $\ket{\psi(t)}$ is written (returning to dimensionful quantities) as an expansion in the number of transitions the system makes between instantaneous eigenstates:
\beq
\ket{\psi(t)}=\ket{\psi(t)}^{(0)}+\ket{\psi(t)}^{(1)}+\ket{\psi(t)}^{(2)}+\cdots.
\label{two}
\eeq
The zeroeth-order term, corresponding to no jumps, is simply $\ket{\psi(t)}^{(0)}=\ket{\psi_{ad}(t)}$, while the first correction is
\bea
\ket{\psi(t)}^{(1)}&=&\sum_{m\neq0}\ket{m(t)}\int_0^t dt_1\,
e^{-i\int_{t_1}^t dt'E_m(t')}\times\nonumber\\
&&\bra{\dot m(t_1)}0(t_1)\rangle
e^{-i\int_{0}^{t_1} dt'\, E_0(t')},
\label{three}
\eea
corresponding to a transition from instantaneous eigenstate $\ket{0}$ to $\ket{m}$ at an arbitrary time $t_1$, sandwiched by adiabatic evolutions.  Subsequent terms can be written by a straightforward generalization of \eqref{three}.

When will the adiabatic approximation be good? At the very least, $||\ket{\psi}^{(1)}||$ must be small. This is of course true for $t$ sufficiently small, but we are often interested in $t=\tau$, in which case generically $\left| \bra{\dot m(t)}0(t)\rangle \right| \sim 1/\tau$, and $||\ket{\psi}^{(1)}||\sim 1$. For a slowly-varying Hamiltonian, this crude estimate neglects the fact that the rapid variation of the phase in the integral in \eqref{three} leads to cancellations in the integral, and indeed it can be shown that \cite{mackenzie-2006-73}
\beq
\left| \langle m(t) \ket{\psi(t)}^{(1)} \right| \sim \left| {\bra{\dot m(t)}0(t)\rangle\over E_m-E_0}\right|.
\label{four}
\eeq
If these terms are all small (and if they are finite in number) then $||\ket{\psi}^{(1)}||$ will be small and, at least to first order, the {\ada} is accurate.

It is perhaps unfortunate that Messiah states that
\beq
\left| {\bra{\dot m(t)}0(t)\rangle\over E_m-E_0}\right| \ll 1
\label{apri}
\eeq
``may be taken as a criterion for the validity of the adiabatic approximation,'' even going so far as to say that it is ``too restrictive" when in fact it can be too loose. (To be fair, he preceded this statement with the qualifier ``in most cases," so he can't be held accountable!) Eq.~\eqref{apri} is often referred to as the ``standard" condition of validity of the {\ada}, but we will refer to it as the {\apri} condition of validity because it can be evaluated without solving for $\ket{\psi(t)}$. At issue is the sufficiency  of \eqref{apri} as a guarantee that the {\ada} $\ket{\psi(t)}\simeq \ket{\psi_{ad}(t)}$ is valid. This can be written in the slightly more convenient form
\beq
1-\left|\bra{0(t)}\psi(t)\rangle\right|\ll 1.
\label{apos}
\eeq
We will refer to \eqref{apos} as the {\apos} condition, because it can be checked only after determining $\ket{\psi(t)}$.

Below, we will argue that \eqref{apri} is a reliable criterion of validity of the {\ada} only insofar as $\left|\bra{\dot m(t)}0(t)\rangle\right|$ is a measure of the time scale of evolution of $H$ (in which case it is $\sim \tau^{-1}$), leading to the widely-held (and generally correct) view that for adiabatic evolution the Hamiltonian must evolve slowly: $E\tau\gg1$. However, if $\left|\bra{\dot m(t)}0(t)\rangle\right|$ is not a measure of the time scale of evolution of the Hamiltonian, then \eqref{apri} cannot be relied upon as a criterion of validity.

The possibility that $\left|\bra{\dot m(t)}0(t)\rangle\right|$ might {\em not\/} be a measure of the time scale of evolution of $H$ can easily be demonstrated with an example. Leaving aside for the moment how quickly $H$ changes, suppose its instantaneous eigenstates do not change appreciably over the course of the evolution, so that $1-\left|\bra{m(t_1)}m(t_2)\rangle\right|\sim\delta\ll1$ for all $t_1,t_2$. Then if the time scale of the evolution of $H$ is $\tau$, we find $\left|\bra{\dot m(t)}0(t)\rangle\right|\sim\delta/\tau$. This being so,
\beq
\left| {\bra{\dot m(t)}0(t)\rangle\over E_m-E_0}\right| \simeq {\delta\over E\tau},
\label{five}
\eeq
and obviously this can be small not because $E\tau\gg1$, but simply because $\delta$ is sufficiently small. Thus, {\em the inadequacy of \eqref{apri} as a criterion of validity for the {\ada} stems from the fact that $|\bra{\dot m(t)}0(t)\rangle|$ may be small even though the Hamiltonian is not slowly varying.} This is our main conclusion.
\medskip

\noindent {\em Spin in uniformly rotating magnetic field}~~~~~A concrete realization of this is the much-studied example of a spin 1/2 in a uniformly-rotating magnetic field, one of very few soluble time-dependent models, making it an obvious testing ground for the sufficiency of the {\apri} criterion. This example was discussed in the context of the {\ada} by Tong et al. \cite{Tong:2005} (inspired by Marzlin and Sanders \cite{Marzlin:2004}, who presented a similar example), and we will comment upon their work shortly. The Hamiltonian for this system can be written
\beq
H(t)=-{\om_0\over2}\bfsi \cdot \bfn(t),
\label{Hamiltonian}
\eeq
where $\bfn(t)=(s_{\theta} c_{\om t},s_{\theta} s_{\om t},c_{\theta})$ (using the shorthand notation $s_x=\sin x$, etc) is the direction of the magnetic field, $\om_0$ is the energy gap between instantaneous eigenstates, $\om$ is the angular frequency of rotation of $\bfn$, and $\theta$ is the opening angle of the cone swept out by $\bfn$. We can (and will) assume $\theta\leq\pi/2$ and $\om_0>0$, but having done so, the sign of $\om$ (the direction of rotation of the magnetic field) cannot be changed without loss of generality, a seemingly innocuous fact that will turn out to be of some importance.

Before proceeding further, note that, intuitively speaking, \eqref{Hamiltonian} is adiabatic -- the Hamiltonian varies slowly -- if $|\om/\om_0|\ll1$.

Let us suppose that the spin is initially aligned with the magnetic field. Then the {\apri} criterion turns out to be \cite{Tong:2005}
\beq
\left| {\om s_{\theta}\over\om_0}\right| \ll 1.
\label{apria}
\eeq
That this does not agree with the intuitive notion of adiabaticity should already cast doubt on whether or not this criterion is sufficient. There is a very simple reason for this discrepancy, which occurs if $\theta\ll1$. If this is so, then the magnetic field is always aligned near the $z$ axis and the instantaneous eigenstates change very little (of order $\theta$). Indeed, $|\langle \pm | \dot\mp\rangle| \sim |\om|\sin\theta$, as envisaged above (see the discussion preceding \eqref{five}).

\begin{figure}[ht]
\begin{center}
\includegraphics[width=4cm]{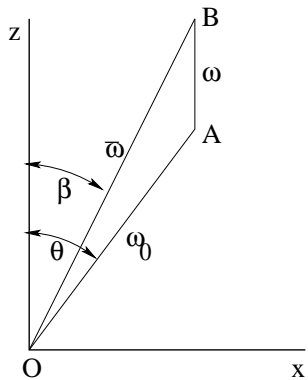}
\end{center}
\caption{Triangle which defines $\bar\omega$ and $\beta$.}
\label{figone}
\end{figure}

The {\apos} criterion can be written
\beq
\left| {\om s_{\theta}\over\bar\om}\right| \ll 1,
\label{aposa}
\eeq
where $\bar\om$ is defined in Fig.~\ref{figone}. (A factor $\sin{\bar\om t/2}$ has been dropped, making \eqref{aposa} the worst-case scenario.) If $\om>0$ then $\bar\om\geq\om_0$, so if \eqref{apria} is satisfied then \eqref{aposa} is as well, which seems to indicate (contrary to the discussion above) that \eqref{apria} is after all a sufficient criterion. However, if $\om<0$ then $\bar\om<\om_0$, and \eqref{apria} being satisfied no longer implies that \eqref{aposa} is also satisfied.

Fig.~\ref{figone} has a physical interpretation familiar from magnetic resonance, which enables us to visualize these conclusions. A common approach to studying a spin in a rotating magnetic field (whether classically or quantum mechanically) is to go into a rotating reference frame, giving a static Hamiltonian. The effect of the change of reference frame is to add a component of the magnetic field along the axis of rotation. Thus, up to a multiplicative constant, the vector OA is the original magnetic field as viewed in the rotating frame, and OB is the effective static magnetic field. The spin, initially in the direction OA, precesses around OB. The instantaneous eigenstate in the rotating frame remains in the direction OA, so the adiabaticity of the evolution depends on the angular separation between OA rotated about OB, on the one hand, and OA itself, on the other. This depends on the angle $\theta-\beta$, and indeed \eqref{aposa} can be written $|\sin(\theta-\beta)|\ll1$. Let us suppose $\theta$ is small, since this is when \eqref{apria} differs significantly from the intuitive notion of adiabaticity. If $\om>0$, then clearly $|\theta-\beta|$ is small and \eqref{aposa} is also satisfied. If, however, $\om<0$, then $|\theta-\beta|$ is not necessarily small and \eqref{aposa} is not automatically satisfied.

To recapitulate, the intuitive notion of adiabaticity for this system is $\om\ll \om_0$. The {\apri} condition differs significantly from this when $\theta\ll 1$, so its appropriateness as a criterion for adiabaticity is already in doubt. This is borne out when $\omega$ is negative and not small. The worst-case scenario is $\omega\simeq -\om_0$. This is in the vicinity of the resonant frequency -- clearly not an adiabatic situation, despite the fact that \eqref{apri} is satisfied. In fact, what is surprising is not so much the fact that \eqref{apri} fails there, but the fact that the evolution actually turns out to be adiabatic if $\om$ is large and positive -- clearly not a situation which satisfies the intuitive notion of adiabaticity.

Although at first sight it is perhaps surprising that the direction of rotation of the magnetic field is important, this is familiar already from the classical problem and is related to the fact that an external magnetic field breaks time reversal and parity invariance. Also, note that the sign of $\omega$ is of no importance to magnetic resonance experiments, because the rotating field is actually an oscillating field, which can be written as a combination of fields rotating in both directions.

How does this relate to the work of Tong, et al.~\cite{Tong:2005}? Let us briefly review their findings. They considered two systems for which the {\apri} conditions agree by construction but the {\apos} ones don't agree, leading to the possibility that both systems satisfy the former condition but only one satisfies the latter. The relation between the two Hamiltonians is $H'(t)=-U^\dagger(t) H(t) U(t)$, where $U(t)$ is the evolution operator for $H(t)$.

They substantiated this idea using \eqref{Hamiltonian} as the unprimed Hamiltonian. The {\apri} criterion \eqref{apria} for $H$ was given, but the {\apos} criterion for $H$ was not in fact evaluated, so the importance of the sign of $\om$, and the inadequacy of \eqref{apria} as a guarantee for adiabatic evolution without even going to the primed system, was not noted.

By construction, the {\apri} condition for the primed system is \eqref{apria}; the {\apos} condition turns out to be \cite{Tong:2005}
\beq
 \sin\theta\ll1.
 \label{aposprime}
 \eeq
The former is satisfied but the latter is not if $\om\ll\om_0$ and $\theta\sim1$, illustrating the inadequacy of \eqref{apri} as a criterion for adiabaticity.

A short calculation shows that the primed Hamiltonian (which was not given in \cite{Tong:2005}) is again a spin 1/2 in a rotating magnetic field, identical in form to \eqref{Hamiltonian} up to a global rotation. We need only make the replacements $\theta\to\theta'=\theta-\beta$ and $\om\to\om'=-\bar\om$. It is important to note that $\om'$ is negative, and also that it is not small in magnitude compared to $\om_0$ if $\om\ll\om_0$ (a fact also noted in Ref.~\cite{Duki:2006}). Noting that $H'$ is of the same form as $H$, the same analysis can be applied simply by making the appropriate substitutions, and nothing new is learned since we have already determined that the {\apri} criterion is inadequate. A primed version of Fig.~\ref{figone} can be used to visualize these conclusions.

While this paper was being written up, a new article has appeared \cite{Tong:2007} which supplements the {\apri} condition with two other conditions which set a limit on the total time of evolution of the system (distinct from the time scale of the Hamiltonian's evolution). This emphasizes a perhaps under-appreciated point: that even for a fixed, slow time scale for the evolution of the Hamiltonian, the system will as a rule eventually escape from the adiabatic state, a fact that has also been observed in \cite{mackenzie-2006-73}. The additional conditions given in \cite{Tong:2007} can be overly restrictive, and even unnecessary, as can be seen by the example \eqref{Hamiltonian} in cases where both \eqref{apria} and \eqref{aposa} are satisfied, since then the {\ada} is valid for all times. This being said, the new conditions (along with \eqref{apri}) are sufficient to guarantee the validity of the {\ada}, which is exactly what the authors set out to do.
\medskip

\noindent {\em Conclusions}~~~~~In  this paper we have emphasized the distinction between the {\adt} and the {\ada}, and have attempted to identify the underlying reason for the inadequacy for what has been labeled the ``standard" criterion (herein called the {\apri} criterion) for  the validity of the {\ada}, \eqref{apri}.  This inadequacy is by no means a weakness of the {\adt}; rather, it is an inadequacy of \eqref{apri} as a measure of the slowness of the time scale of evolution of the Hamiltonian -- the intuitive notion of adiabaticity. While often a reliable indicator of slowness of the Hamiltonian, \eqref{apri} can fail to do so in situations where the Hamiltonian changes quickly but not by a great amount. In these situations the {\ada} can be invalid despite the fact that \eqref{apri} is satisfied. This point of view has been illustrated by studying a spin 1/2 in a rotating magnetic field.

Finally, we have emphasized that even for a slowly-evolving Hamiltonian, the system will not stay in the adiabatic state indefinitely; the system will as a rule eventually escape, although this may take a time much longer than the actual time scale on which the Hamiltonian evolves. This point has been emphasized in \cite{mackenzie-2006-73}, and has been studied more quantitatively by others more recently \cite{Tong:2007}.
\medskip

\noindent {\em Acknowledgments}~~~~~This work was funded in part by the
National Science and Engineering Research Council of Canada.

\bibliographystyle{unsrt}
\bibliography{fieldtheory}

\begin{thebibliography}{10}

\bibitem{Born:1928}
M.~Born and V.~Fock.
\newblock Beweis des adiabatensatzes.
\newblock {\em Z. Phys.}, 51:165--169, 1928.

\bibitem{Tong:2005}
D.M. Tong, K.~Singh, L.C. Kwek, and C.H. Oh.
\newblock Quantitative conditions do not guarantee the validity of the
  adiabatic approximation.
\newblock {\em Phys. Rev. Lett.}, 95:110407, 2005.

\bibitem{Marzlin:2004}
Karl-Peter Marzlin and Barry~C. Sanders.
\newblock Inconsistency in the application of the adiabatic theorem.
\newblock {\em Phys. Rev. Lett.}, 93:160408, 2004.

\bibitem{Duki:2006}
Solomon Duki, H.~Mathur, and Onuttom Narayan.
\newblock Is the adiabatic approximation inconsistent?
\newblock 2006.

\bibitem{Sarandy:2004}
M.S. Sarandy, L.-A. Wu, and D.A. Lidar.
\newblock Consistency of the adiabatic theorem.
\newblock {\em Quant. Inform. Proc.}, 3:331, 2004.

\bibitem{Pati:2004}
A.K. Pati and A.K. Rajagopal.
\newblock Inconsistencies of the adiabatic theorem and the {Berry} phase.
\newblock {\em quant-ph/0405129}, 2004.

\bibitem{Tong:2004}
D.M. Tong, K.~Singh, L.C. Kwek, and C.H. Oh.
\newblock A note on the geometric phase in adiabatic approximation.
\newblock {\em Phys. Lett. A}, 339:288, 2005.

\bibitem{Ambainis:2004}
Andris Ambainis and Oded Regev.
\newblock An elementary proof of the quantum adiabatic theorem.
\newblock {\em quant-ph/0411152}, 2004.

\bibitem{Wu:2004}
Zhaoyan Wu, Li~Zheng, and Hui Yang.
\newblock On ``inconsistency in the application of the adiabatic theorem''.
\newblock {\em quant-ph/0411212}, 2004.

\bibitem{Ye:2005}
Ming-Yong Ye, Xiang-Fa Zhou, Yong-Sheng Zhang, and Guang-Can Guo.
\newblock Condition for the adiabatic approximation.
\newblock {\em quant-ph/0509083}, 2005.

\bibitem{Vertesi:2005}
T.~V\'ertesi and R.~Englman.
\newblock Perturbative analysis of possible failures in the traditional
  adiabatic conditions.
\newblock {\em quant-ph/0511141}, 2005.

\bibitem{Ma:2006}
Jie Ma, Yongping Zhang, Enge Wang, and Biao Wu.
\newblock Comment on inconsistency in the application of the adiabatic theorem.
\newblock {\em Phys. Rev. Lett.}, 97:128902, 2006.

\bibitem{Messiah:1962}
Albert Messiah.
\newblock {\em Quantum Mechanics, vols. 1 and 2}.
\newblock Wiley, 1962.

\bibitem{mackenzie-2006-73}
R.~MacKenzie, E.~Marcotte, and H.~Paquette.
\newblock A new perturbative approach to the adiabatic approximation.
\newblock {\em Physical Review A}, 73:042104, 2006.

\bibitem{Tong:2007}
D.M. Tong, K.~Singh, L.C. Kwek, and C.H. Oh.
\newblock Sufficiency criterion for the validity of the adiabatic
  approximation.
\newblock {\em Phys. Rev. Lett.}, 98:150402, 2007.

\end{thebibliography}
\end{document}